\documentclass[aps,prd,twocolumn,groupedaddress]{revtex4}
\usepackage{amsmath}

\begin{document}

\newcommand{\comment}[1]{}
\renewcommand{\textit}[1]{}

\title{The singular field used to calculate the self-force on non-spinning and
spinning particles}

\author{Eirini Messaritaki}
\email[Email: ]{emess@caltech.edu}
\altaffiliation{Current address: California Institute of
Technology, MS 18-34, Pasadena, CA 91125, USA}
\affiliation{Department of Physics, P.O. Box 118440, University of Florida,
Gainesville, Florida 32611-8440, USA}
\affiliation{Department of Physics, P.O. Box 413, University
of Wisconsin - Milwaukee, Milwaukee, Wisconsin, 53201, USA}

\date{\today}

\begin{abstract}

The singular field of a point charge has recently been described
in terms of a new Green's function of curved spacetime. This singular
field plays an important role in the calculation of the self-force acting upon
the particle. We provide a method for calculating the singular field
and a catalog of expansions of the singular
field associated with the geodesic motion of monopole and
dipole sources for scalar, electromagnetic and gravitational fields. These
results can be used, for example, to calculate the effects of the
self-force acting on a particle as it moves through spacetime.
\end{abstract}


\maketitle

\section{Introduction}
\label{Intro}

Detailed knowledge of the evolution of a binary system consisting
of a stationary supermassive black hole (of mass of the order of
$10^6 M_\odot$) and a stellar-mass neutron star
or black hole would be an important aid for the analysis of data
from space-based gravitational wave detectors such as LISA
\cite{LISAreference}. In general, the stellar-mass neutron star or black hole,
usually modelled as a pointlike particle, inspirals toward the
central black hole due to radiation reaction. Radiation reaction
is only one consequence of the self-force on a
particle, namely the force which results from the interaction of the
particle with its own gravitational field.

Generally, the spin of the stellar-mass compact object is
expected to have an effect on the gravitational waveforms by such
a system. 
That effect is expected to be small and does not need to be taken into
account in template waveforms aimed at detection of such binaries.
However it will have to be taken into account 
for accurate estimation of the physical parameters of the
binary system. For that reason, self-force
calculations for combined monopole and dipole sources are required.
This paper is a step toward self-force calculations for spinning 
sources.

\subsection{Self-Force Calculation}

One method for calculating the self-force on a particle generating an 
electromagnetic field was suggested by DeWitt and Brehme \cite{dewbre}.
That method involves decomposing the retarded field at point $p$,
generated by the charged particle at point $p'$, into its direct part
(which comes from the part of the Green's function with
support only on the null cone of point $p$) and its
tail part (which comes from the part of the Green's function with
support only within the null cone of point
$p$), and then using the tail part of the electromagnetic field in
the Lorentz force law.
Similar analysis by
Mino, Sasaki and Tanaka \cite{mst} described the
gravitational self-force on a massive particle in terms
of the tail part of the gravitational field. Axiomatic approaches
were used by Quinn for the
scalar field \cite{q00} and by Quinn and Wald for the electromagnetic and
gravitational fields \cite{qw97} and arrived at similar
conclusions.

The actual calculation of the self-force on a particle in geodesic
motion around a Schwarzschild black hole, in terms of the tail
part of the field, has been successfully performed by a variety
of groups
\cite{mns,bo0,bo2,bo1,burko,BarBur,barack1,barack2,BMetal,BarackLousto02}.
In these applications, the retarded field is first calculated via
separation of variables coupled with a sum over modes.
Analytic methods are used to calculate the part of the self-force
from each mode of the direct part of the field.  Then, for each
mode, the part of the force from the direct part is subtracted from the
part of the force
from the retarded part. Finally, this difference is summed over
all modes and this sum provides the self-force
which results from the tail part of the field. 
It is clear, then, that the tail part
provides a useful calculational avenue to the self-force. However, a
few shortcomings remain on a fundamental level \cite{detwhi02}.
The Green's function of the tail part has support inside the
past null cone of the field point, which implies that this
description of the self-force formally depends upon the entire
past history of the particle. Also, in some circumstances the tail
part is finite but not differentiable at the location of the
particle and averaging of the derivative over different
directions of approach to the particle is required to determine
the actual self-force. Perhaps most critically, the tail part of
the field is not a solution of the field equations. Thus,
the tail part of the field of a particle
can be described mathematically, but it cannot by itself be considered a
field.

\subsubsection{Self-force calculation using the singular field}

A different method for calculating the self-force was proposed by
Detweiler and Whiting \cite{detwhi02} for the scalar, electromagnetic and
gravitational fields. That method uses a new Green's function for
curved spacetime to determine the {\it singular} field of the particle.
The singular field was first defined in \cite{detwhi02} where it was shown
by construction to be the part of the retarded field of the particle that 
exerts no force on the particle itself. The properties of the singular
field were detailed in section 13.5 of \cite{PoissonsReview}. 
For completeness, we give here the definition and properties of the singular 
field. For a more detailed discussion, including the motivation for the 
definition of the singular field, we point the reader to \cite{detwhi02} and 
\cite{PoissonsReview}.

Assume a particle of scalar charge $q$ at point $p'$ in a background
spacetime. We want to calculate the retarded field of that charge at point $p$ 
in spacetime. The retarded Green's function $G^\text{ret}$
obeys the inhomogeneous equation
\begin{equation}
\nabla^2 G^\text{ret}(p,p') = - 4 \pi \delta(p,p')
\label{eq:retarded}
\end{equation}
where $\nabla$ denotes differentiation with respect to the background metric.
The retarded Green's function generates a field which is the physically relevant
solution, because it obeys appropriate outgoing-wave boundary conditions at 
infinity.

The singular Green's function $G^\text{S}$ is defined 
\cite{detwhi02, PoissonsReview} as the function
that satisfies the same inhomogeneous equation as the retarded Green's
function
\begin{equation}
\nabla^2 G^\text{S}(p,p') = - 4 \pi \delta(p,p'),
\label{eq:singular}
\end{equation}
with the additional requirements that it is symmetric in its arguments $p$ 
and $p'$ and that it vanishes if $p$ is within the past or future null cones of 
$p'$. 
By definition, the singular Green's function has the same singular behavior at 
the location of the particle as the retarded Green's function: the divergence
of the field near the particle is insensitive to the boundary conditions 
at infinity. 
It has been shown that such a function
always exists in the local convex neighborhood of $p'$ (see chapter 13 of
\cite{PoissonsReview}) and that the singular field that results from it  
does not exert any force on the particle itself. 
As explained in \cite{PoissonsReview}, because the singular Green's function
is symmetric in its arguments, it does not distinguish between past and
future and it produces a field that contains equal amounts of outgoing and
incoming radiation at infinity. Removing
$G^\text{S}$ from the retarded Green's function $G^\text{ret}$ has the effect 
of removing the singular behavior of the field without that affecting the 
motion of the particle.

The {\it radiative} Green's function, also referred to as the {\it regular
remainder}, is defined by
\begin{equation}
G^\text{R}(p,p') = G^\text{ret}(p,p') - G^\text{S}(p,p').
\label{eq:radiative}
\end{equation}
This definition implies that the radiative Green's function obeys the 
homogeneous equation
\begin{equation}
\nabla^2 G^\text{R}(p,p') = 0, 
\end{equation}
that it is equal to the retarded Green's function if $p$ is within the 
future null cone of $p'$ and that it vanishes if $p$ is in the 
past null cone of $p'$ \cite{PoissonsReview}. The radiative Green's 
function produces a field that is smooth and differentiable on the
world line of the particle and which properly encodes the outgoing-wave
boundary conditions at infinity.

As has already been mentioned, after the singular field is subtracted from 
the retarded field, the radiative field is entirely responsible for the 
self-force. Specifically for the scalar case
\begin{equation}
\psi^{\text{R}}(p) = \psi^{\text{ret}}(p) - \psi^{\text{S}}(p)
\label{BreakUp}
\end{equation}
and the self-force is calculated by using the radiative field
in the self-force equation
\begin{equation}
\mathcal{F}^{a} = q \lim_{p \to p'} \nabla^a \psi^{\text{R}}(p) = q \lim_{p
\to p'} \nabla^a [\psi^{\text{ret}}(p) - \psi^{\text{S}}(p) ].
\end{equation}
As was mentioned in \cite{detwhi02}, for the self-force calculation
for the electromagnetic case knowledge of the singular electromagnetic
potential is required, while for the self-force calculation for
the gravitational case knowledge of the singular gravitational field
is required. The corresponding equations are given in 
\cite{detwhi02}.

We argue here that the method for calculating the self-force that uses
the singular and radiative fields is preferable to the one that uses the 
direct and tail fields.
For the scalar, electromagnetic and gravitational cases, the singular field 
obeys the inhomogeneous Poisson, Maxwell and Einstein equations respectively 
and has support only outside the null cone of the particle.
In that sense, knowledge of the entire past history of the particle is not
required, as is the case for the tail field, and the singular
field needs to be calculated locally in the neighborhood of the
point particle. Additionally,
the radiative field obeys the corresponding homogeneous
equation and thus is finite and differentiable everywhere along
the worldline of the particle. Consequently, this method for calculating
the self-force does not present the difficulties of
interpretation inherent in the method involving the direct and tail
fields. 

A recent self-force calculation using the singular and
radiative fields \cite{detmeswhi} showed that this method gives
identical results to the ones derived using the 
direct and tail fields.
The results were also used \cite{diazmesswhidet} to predict the
self-force effects on various orbits of scalar particles in a
Schwarzschild background. Finally, a different approach to the self-force
calculation described in \cite{KFW2006} uses the singular field in 
order to extract the contribution of that field to the Weyl scalar
$\it{\Psi_0^{\text{S}}}$ (or $\it{\Psi_4^{\text{S}}}$), 
subtract it from $\it{\Psi_0^{\text{ret}}}$ 
($\it{\Psi_4^{\text{ret}}}$) resulting from the retarded field of the particle
and use the renormalized $\it{\Psi_0^{\text{ren}}}$ 
($\it{\Psi_4^{\text{ren}}}$) to calculate the
perturbations on the background spacetime.

\subsection{Outline}
In this paper we establish a method for calculating the singular field
for monopole and dipole sources of scalar, electromagnetic and gravitational 
fields and we present a catalog of the expansions of the
singular field for those sources.
The coordinates that are used to calculate the
expansion of the singular field are the Thorne-Hartle-Zhang coordinates,
(abbreviated as THZ coordinates). Those coordinates were
initially introduced by Thorne and Hartle \cite{TH} and later extended by
Zhang \cite{Z}. A short discussion of them is
presented in Sec.~\ref{THZ}.

Expansions are given for the singular fields associated with all
monopole and dipole sources for scalar and electromagnetic fields
and the monopole gravitational
field in Sec.~\ref{ScalarField}, \ref{ElectromagneticPotential} and
\ref{GravitationalField}, respectively. 
A short discussion of the results is given in Sec.~\ref{Discussion}
and future work is outlined in Sec.~\ref{future}.

\section{THZ Coordinates}
\label{THZ}

The Poisson, Maxwell and Einstein equations for the singular field assume
a relatively simple form when written in a coordinate system in
which the background spacetime looks as flat as possible. In the
following, it is assumed that the particle is moving on a geodesic
$\Gamma$ in a vacuum background described by the metric $g_{ab}$. Also,
$\mathcal{R}$ is a representative length scale of the background geometry,
the smallest of the radius of curvature, the scale of inhomogeneities of
the background and the time scale of curvature changes along the geodesic
$\Gamma$.

For a geodesic $\Gamma$, locally inertial coordinates $x^a = (t,x,y,z)$ can 
be found so that, on the geodesic, the metric and its first derivatives 
coincide with the Minkowski
metric and $t$ measures the proper time along the geodesic \cite{mtw}. 
Such a locally inertial coordinate system is not unique. The THZ
coordinate system used here was first introduced by Thorne and Hartle
\cite{TH} and later extended by Zhang \cite{Z} and has meaning only locally, 
close to the
worldline of the particle. Specifically it is assumed that the background
metric close to the worldline of the particle can be written as
\begin{eqnarray}
g_{ab} &=& \eta_{ab} + H_{ab} \\ \nonumber
    &=& \eta_{ab} + {}_{2}H_{ab}   + {}_{3}H_{ab}
    + O(\frac{\rho^4}{\mathcal{R}^4}), \nonumber
\label{TaylorExpandMetric}
\end{eqnarray}
where $\eta_{ab}$ is the flat Minkowski metric in the THZ coordinates
$(t,x,y,z)$ and
\begin{equation}
\rho^2 = x^2 + y^2 + z^2.
\end{equation}
Also
\begin{equation}
\begin{split}
_{2}H_{ab} &dx^a dx^b = - \mathcal{E}_{ij} x^i x^j ( dt^2 + \delta_{kl}
dx^k dx^l ) \\
&+ \frac{4}{3} \epsilon_{kpq} \mathcal{B}^q
{}_i x^p x^i dt dx^k \\
&- \frac{20}{21} \Big [ \dot{\mathcal{E}}_{ij} x^i x^j x_k - \frac{2}{5} \rho^2
 \dot{\mathcal{E}}_{ik}x^i \Big ] dt dx^k \\
&+ \frac{5}{21} \Big [ x_i \epsilon_{jpq} \dot{\mathcal{B}}^q{}_k x^p x^k -
\frac{1}{5} \rho^2 \epsilon_{pqi} \dot{\mathcal{B}}_j{}^q x^p \Big ] dx^i dx^j,
\end{split}
\label{H2ab}
\end{equation}
\begin{equation}
\begin{split}
_{3}H_{ab} dx^a dx^b = &-\frac{1}{3} \mathcal{E}_{ijk} x^i x^j x^k(dt^2+
\delta_{nl}dx^n dx^l) \\
    &+\frac{2}{3} \epsilon_{kpq} \mathcal{B}^q{}_{ij}x^px^ix^jdtdx^k \\
 &+ O \big (\frac{\rho^4}{\mathcal{R}^4} \big ) _{ij} dx^i dx^j.\\
\end{split}
\label{H3ab}
\end{equation}
In this and the following, $a$, $b$, $c$ and $d$ denote spacetime indices.
The indices $i, j, k, l, n, m, p$ and $q$ are spatial indices and, to the
order up to which the calculations are performed, they are raised and
lowered by the 3-dimensional flat space metric $\delta_{ij}$. The dot
denotes differentiation with respect to the time $t$ along the geodesic.
Also, $\epsilon_{ijk}$ is the 3-dimensional flat space antisymmetric
Levi-Civita tensor.

The tensors $\mathcal{E}$ and $\mathcal{B}$ are spatial, symmetric and
trace-free and their components are related to the Riemann tensor on the
geodesic $\Gamma$ by
\begin{eqnarray}
\mathcal{E}_{ij} &=& R_{titj} \\
\mathcal{B}_{ij} &=& \frac{1}{2} \epsilon_i{}^{pq} R_{pqjt} \\
\mathcal{E}_{ijk} &=& \big [\nabla_k R_{titj} \big]^{\text{STF}} \\
\mathcal{B}_{ijk} &=& \frac{3}{8} \big[\epsilon_i{}^{pq} \nabla_kR_{pqjt}
\big]^{\text{STF}}
\label{EandB}
\end{eqnarray}
where STF means to take the symmetric, trace-free part with
respect to the spatial indices $i$, $j$ and $k$. The components
$\mathcal{E}_{ij}$ and $\mathcal{B}_{ij}$ are of
$O({1}/{\mathcal{R}^2})$ and their time derivatives are of
$O({1}/{\mathcal{R}^3})$. The components $\mathcal{E}_{ijk}$ and
$\mathcal{B}_{ijk}$ are also of $O({1}/{\mathcal{R}^3})$.

If $H_{ab}$ consists of the terms given in Eq.~(\ref{H2ab}), the
coordinates are second-order THZ coordinates and are well defined up to
the addition of arbitrary functions of $O({\rho^4}/{\mathcal{R}^3})$
\cite{Det2005}.
Notice that in that case, only the Riemann tensor components and their
time derivatives appear in the metric.
If $H_{ab}$ also includes the terms given in Eq.~(\ref{H3ab}), the
coordinates are third-order THZ coordinates and are well defined up to the
addition of arbitrary functions of $O({\rho^5}/{\mathcal{R}^4})$ \cite{Det2005}.
In this case, only spatial derivatives of the Riemann tensor components
appear in the metric.

The contributions to the metric components $_2H_{ab}$ that contain 
$\dot{\mathcal{E}}_{ij}$ and $\dot{\mathcal{B}}_{ij}$ are of 
$O(\rho^3 / \mathcal{R}^3)$, as are the contributions to the metric components 
$_3H_{ab}$ that contain ${\mathcal{E}}_{ijk}$ and ${\mathcal{B}}_{ijk}$.
Therefore, any calculation that includes the $\dot{\mathcal{E}}_{ij}$ and
$\dot{\mathcal{B}}_{ij}$ terms of ${}_2 H_{ab}$ must also include
the ${\mathcal{E}}_{ijk}$ and ${\mathcal{B}}_{ijk}$ terms of
${}_3 H_{ab}$ as well.

Using the simple symmetry properties of the tensors $\mathcal{E}$ and
$\mathcal{B}$, the following relationships can be shown for their
components and the spatial THZ coordinates $(x,y,z)$:
\begin{equation}
\begin{split}
&\epsilon_{ijk} \mathcal{E}^k{}_l + \epsilon_{ikl} \mathcal{E}^k{}_j -
\epsilon_{jkl} \mathcal{E}^k{}_i = 0 \\
&\epsilon_{ijk} \mathcal{B}^{k}{}_l + \epsilon_{ikl} \mathcal{B}^k{}_j -
\epsilon_{jkl} \mathcal{B}^k{}_i = 0
\end{split}
\label{EBRel2}
\end{equation}
and
\begin{equation}
\begin{split}
&[\epsilon_{ijk}( \mathcal{E}^k{}_n x_l - \mathcal{E}_{ln} x^k) + \epsilon_{ikl}
 \mathcal{E}^k{}_n x_j - \epsilon_{jkl} \mathcal{E}^k{}_n x_i ] x^n x^l = 0 \\
&[\epsilon_{ijk}( \mathcal{B}^k{}_n x_l - \mathcal{B}_{ln} x^k) + \epsilon_{ikl}
 \mathcal{B}^k{}_n x_j - \epsilon_{jkl} \mathcal{B}^k{}_n x_i ] x^n x^l =
 0.
\end{split}
\label{EBRel4}
\end{equation}
These relationships are used in Sec.~\ref{Calc} where the
singular fields for different sources are calculated, in order 
to simplify the expressions for those fields.

Before we proceed to the calculation of the singular fields, we need to make
an important point on the notation and language that we use.
In the following, the subscripts (or superscripts) (0), (2), (3) and (4)
are used to indicate the order of significance of each term
or component, for the singular fields, the covariant derivatives with respect
to the THZ metric and the components of the Riemann tensor that show
up in the Poisson, Maxwell and Einstein equations below. The subscript 
(0) refers to the most dominant
contribution (resulting from the part of the THZ metric that is of
order 1). The subscripts (2) and (3) refer to the 
next two more significant corrections (resulting from the parts of the
THZ metric that are of order $(\rho^2 / \mathcal{R}^2)$
and $(\rho^3 / \mathcal{R}^3)$ respectively), which are calculated
for the singular fields. The subscript (4) refers to the next
correction (resulting from the $(\rho^4 / \mathcal{R}^4)$ part of the metric). 
The subscript (1) is not used due to the fact that there are no
$(\rho / \mathcal{R})$ terms in the metric.
The terms ``first correction'', ``second correction'' and 
``third correction'' refer to the correction coming from the
$O(\rho^2 / \mathcal{R}^2)$, $O(\rho^3 / \mathcal{R}^3)$ 
and $O(\rho^4 / \mathcal{R}^4)$ parts of
the metric respectively. The fact that they are labeled with indices
(2), (3) and (4) (instead of the intuitive choice (1), (2) and (3), 
respectively) should not cause any confusion, since the reason for
it is justified.

\section{Calculation of the Singular Fields}
\label{Calc}

The singular fields for a variety of sources are calculated below. The
calculations were done using the program \textsc{GrTensor} \cite{grtensor}
running under \textsc{Maple}.

It should be noted that in all cases it is assumed that the characteristic
of the moving source that is responsible for its own 
field (i.e. charge, dipole moment, mass) is 
small compared to the background curvature. That assumption allows
us to safely consider the effects of the self-force as a small
perturbation on the motion of the particle.

The derivation of each 
singular field follows a similar pattern.  To illustrate that
pattern we use the simple example of the scalar monopole field. In
THZ coordinates the monopole scalar field is assumed to have
$q/\rho$ as the leading term in the expansion in powers of the
distance away from the point source.  This singular behavior
correctly accounts for the $\delta$-function source at the origin.
However, the $\nabla^2$ operator in curved spacetime, in THZ
coordinates, differs from its flat space counterpart by terms
involving the $\mathcal{E}_{ij}$ and $\mathcal{B}_{ij}$ multipole
moments; these have dimensions of $1/(\text{length})^2$ and are of
$O(1/\mathcal{R}^2)$.  It is obvious that the first
correction to the singular monopole scalar field must behave as $q
(\mathcal{E}_{ij}\text{ or }\mathcal{B}_{ij}) x^i x^j/\rho$ in order to
have the proper index structure and dimension. The correction must
have no free indices, and the two indices on either of the the
tracefree $\mathcal{E}_{ij}$ or $\mathcal{B}_{ij}$ require
precisely the $x^i$ and $x^j$. The power of $\rho$ in the
denominator is determined by the need to have the correct overall
dimension. With a similar argument it follows that the succeeding
correction to the singular monopole field 
should generally be of the form 
$q(\mathcal{E}_{ijk}\text{ or }\mathcal{B}_{ijk}) x^i x^j x^k/\rho$.

This picture of the expansion is consistent with the general
description of the Hadamard expansion \cite{hadamard}
of the Green's function in curved
spacetime for the singular field \cite{dewbre,detwhi02}.
The issue that remains is to verify that
this expansion does not surreptitiously include some of what
ought to be considered the ``regular remainder'' R-part of the 
retarded filed, which is responsible
for the self-force.  Assume, for the moment,
that the expansion of the correct singular monopole field does include a regular
piece; such a regular piece must be a homogeneous solution of the field
equations in the vicinity of the particle because the $\delta$-function
source is already accounted for in the expansion which was just described.
Thus, the general form of such a correction must be expandable in THZ
coordinates about the particle, and must behave as
\begin{equation}
  \text{constant} + \text{linear term} + \text{quadratic} + \text{cubic} +
\ldots .
\end{equation}
The first derivative of the constant term vanishes at the origin, and
therefore could exert no force on the particle.  Similarly the first
derivatives of the quadratic, cubic and higher order terms also vanish
at the location of the particle and cannot contribute to the self-force.
Only the first derivative of the linear term remains. However, the linear term
ought to be describable only in terms of the geometry and should be
expressible as 
\begin{equation}
  q (\mathcal{E}_{ij}\text{ or }\mathcal{B}_{ij}) \times
      (\text{something linear in } x^i) .
\end{equation}
But there is no ``something linear'' in $x^i$ which has the proper
dimensions and index structure.  
Thus, while the singular monopole field constructed above
might, in principle, include an extra ``homogeneous part,'' the
extra part could not be linear in $x^i$ and could not affect the
actual self-force on the particle. A similar argument can be used
to rule out the presence of at least regular quadratic and cubic
terms as well.

It will be seen in the following that the order up to which we
have to calculate the singular fields (in order to be able to
calculate the self-force) depends on the particle in question. In
some cases, it is sufficient to calculate the first correction, namely
that which
results from the $O(\rho^2 / \mathcal{R}^2)$ parts of the THZ
metric. In the other cases, we must also know the second correction
that comes from the $O(\rho^3 / \mathcal{R}^3)$ 
parts of the metric. For completeness, we calculate
all these corrections for all the sources that we examine and we
explicitly state in which cases that is not strictly necessary. 

In the cases in which calculating the second correction is not
strictly necessary, we can still significantly benefit from such a
calculation. Firstly, the more accurate the approximation of the
singular field, the more differentiable the radiative field.
For example, as explained in \cite{detmeswhi}, if the approximation to 
the scalar singular field is in error by a $C^\text{n}$ function, the 
approximation
to the scalar radiative field is no more differentiable than $C^\text{n}$ and 
the
approximation to the self-force (obtained by differentiating the
radiative field) is no more differentiable than $C^\text{n-1}$. Additionally,
according to the standard method for calculating the
self-force \cite{bo0},one needs to do a spherical harmonic decomposition
for the retarded and singular fields, subtract the l-mode contribution
of each one and finally sum over all l-modes.
As was seen in \cite{detmeswhi} and in \cite{mythesis},
the convergence of the sum over the spherical harmonics can be
sped up by including the spherical harmonic decomposition of 
higher order corrections.

\subsection{Scalar Field}
\label{ScalarField} Assume that a particle moving on a
background geodesic creates a scalar source $\varrho$. The scalar
field generated by the particle can be expanded as
\begin{equation}
\Psi^{\text{S}} = \Psi^{\text{S}}_{(0)} + \Psi^{\text{S}}_{(2)} +
\Psi^{\text{S}}_{(3)} + \ldots
\end{equation}
and it obeys Poisson's equation
\begin{equation}
\nabla^{2} \Psi^{\text{S}} = \nabla_{(0+2+3+\ldots)}^{2} (
\Psi^{\text{S}}_{(0)} + \Psi^{\text{S}}_{(2)} +
\Psi^{\text{S}}_{(3)} +\ldots) = -4 \pi \varrho. \label{GeneralEq}
\end{equation}
Regardless of the form of $\varrho$, the leading term in the
expansion is the
scalar field generated by the particle when it is stationary at
the origin of a Cartesian coordinate system and it obeys the
lowest order differential equation derived from Eq.~(\ref{GeneralEq})
\begin{equation}
\nabla_{(0)}^{2} \Psi^{\text{S}}_{(0)} = -4 \pi \varrho.
\end{equation}

The first correction obeys the differential equation derived from
Eq.~(\ref{GeneralEq}) by keeping only the lowest-order terms, namely
\begin{equation}
\nabla_{(0)}^{2} \Psi^{\text{S}}_{(2)} + \nabla_{(2)}^{2}
\Psi^{\text{S}}_{(0)} = 0 
\label{Scalar1}
\end{equation}
which means that the $\nabla^{2}_{(2)}$ acting on the zeroth-order
part of the field is equal to minus the source term in the scalar differential
equation for the first correction to the field. 
In general, the source term in this equation is expected to
contain the components of the tensors $\mathcal{E}$ and
$\mathcal{B}$ and to give the first correction coming from
the particle's motion on the background geodesic.
From this equation it can be inferred that 
\begin{equation}
\Psi^{\text{S}}_{(2)} \sim 
[O(\Psi^{\text{S}}_{(0)})\times O(\frac{\rho^2}{\mathcal{R}^2})].
\label{ScalarOrder1}
\end{equation}

The second correction obeys the differential equation derived from
Eq.~(\ref{GeneralEq}) by keeping the $O({\rho^3}/{\mathcal{R}^3})$ terms
of the metric, namely
\begin{equation}
\nabla_{(0)}^{2} \Psi_{(3)}^{\text{S}} + \nabla_{(3)}^{2}
\Psi_{(0)}^{\text{S}} =0. \label{Scalar2}
\end{equation}
The source term in this case is identically equal to minus the
$O({\rho^3}/{\mathcal{R}^3})$-part of $\nabla^2$ acting on
$\Psi_{(0)}^{\text{S}}$ and in principle involves
 $\mathcal{E}_{ijk}$ and $\mathcal{B}_{ijk}$ as well as the
time-derivatives of $\mathcal{E}_{ij}$ and $\mathcal{B}_{ij}$.
From this equation it can be inferred that 
\begin{equation}
\Psi^{\text{S}}_{(3)} \sim 
[O(\Psi^{\text{S}}_{(0)})\times O(\frac{\rho^3}{\mathcal{R}^3})].
\label{ScalarOrder2}
\end{equation}
Notice that $\Psi_{(2)}^{\text{S}}$ is not included in the equation for
$\Psi^{\text{S}}_{(3)}$, because it shows up in the term $\nabla_{(2)}^2
\Psi_{(2)}^{\text{S}}$, a term of order $[O(\Psi^{\text{S}}_{(0)})
\times O({\rho^4}/{\mathcal{R}^4})]$, which must be included
in the calculation of the third correction. 

The source for the third correction contains both
$\Psi^{\text{S}}_{(0)}$  and $\Psi^{\text{S}}_{(2)}$.
Specifically, the differential equation derived from
Eq.~(\ref{GeneralEq}) by keeping all metric terms of 
$O({\rho^4}/{\mathcal{R}^4})$ is
\begin{equation}
\nabla_{(0)}^2 \Psi_{(4)}^{\text{S}} + \nabla_{(2)}^2
\Psi_{(2)}^{\text{S}} + \nabla_{(4)}^2 \Psi_{(0)}^{\text{S}} = 0
\end{equation}
in which the second and third terms are effectively the source terms.
That gives that the third correction is
\begin{equation}
\label{OrderScalar}
\Psi_{(4)}^{\text{S}} \sim
[O(\Psi_{(0)}^{\text{S}}) \times
O(\frac{\rho^4}{\mathcal{R}^4})].
\end{equation}

\subsubsection{Monopole Field}
Assume that the particle in question is carrying a scalar charge $q$
and is moving on a geodesic.
The scalar field generated by that particle is the scalar
monopole field. The detailed calculation of that field in the THZ
coordinates was
presented in \cite{detmeswhi} and the result is given here for
completeness.

The zeroth-order term is the Coulomb-like potential
$(q/\rho)$. Both the first and second order corrections can be shown
to be equal to 0 due to symmetries, so the result is
\begin{equation}
\Psi^{\text{S}} = \frac{q}{\rho} + O(\frac{\rho^3}{\mathcal{R}^4}).
\end{equation}

Disregarding the fact that the first and second
corrections to the field are equal to zero, Eq.~(\ref{ScalarOrder1})
and (\ref{ScalarOrder2}) imply that one would only need to calculate
up to the first correction of the scalar monopole singular field
in order to be able to 
calculate the self-force on the scalar charge. That is because the
second correction, were it not zero, would be of $O(\rho^2 /
\mathcal{R}^3)$ which would give no contribution 
to the self-force on the particle after the first derivative
followed by the limit to the location of the particle were taken.

\subsubsection{Dipole Field}
\label{SFD}

Assume that the particle is carrying a scalar dipole moment and is
moving on a geodesic of the given background. The dipole moment is
assumed to have a random orientation and its THZ components
 are denoted as $K_{a} = (0, K_x, K_y, K_z)$.

The leading term in the scalar field series expansion is the scalar
field generated by a dipole that is stationary at the origin of a
Cartesian coordinate system and is 
\begin{equation}
\Psi^{\text{S}}_{(0)} = \frac{K_i x^i}{\rho^3}.
\end{equation}

For the first correction, explicit evaluation of Eq.~(\ref{Scalar1}) 
shows that $\nabla^2_{(2)} \Psi^{\text{S}}_{(0)}=0$, which results in
\begin{equation}
\big ( \frac{\partial^2}{\partial x^2} + \frac{\partial^2}{\partial y^2} +
\frac{\partial^2}{\partial z^2} \big ) \Psi_{(2)}^{\text{S}} = 0,
\end{equation}
so the first correction to the field can be set equal to a
constant, or, for self-force calculations, equal to 0.

Eq.~(\ref{Scalar2}) gives that the second correction to the singular
field obeys a differential equation that relates the second derivatives of
$\Psi^{\text{S}}_{(3)}$ to terms of the form
\begin{displaymath}
K_. \dot{\mathcal{B}}_{..} \frac{x^. x^. x^. x^. x^. x^.}{\rho^7}
\end{displaymath}
where the dots denote appropriately contracted indices for each term.
The tensors $\mathcal{E}_{ijk}$ or $\mathcal{B}_{ijk}$ do not show
up in the equation.
The fact that there is a large number of terms in the differential equation
makes it very inconvenient to present the equation explicitly here. In order
to give the reader an idea of what the equation looks like, we explicitly 
write a few terms of it
\footnote{The differential equations become more complicated as the
electromagnetic potentials and the gravitational fields are
considered. For that reason, the differential equation is only 
shown for this simpler case of the scalar dipole field. It should be
kept in mind that it has the same general form for the more complicated
fields as well.}, namely
\begin{equation}
\big ( \frac{\partial^2}{\partial x^2} + \frac{\partial^2}{\partial y^2} +
\frac{\partial^2}{\partial z^2} \big ) \Psi_{(3)}^{\text{S}} +
K_x \dot{\mathcal{B}}_{yz} \frac{x^2 y^2 z^2}{\rho^7} + \hdots = 0.
\end{equation}
Using arguments similar to the ones mentioned at the beginning of
Sec.~(\ref{Calc}) for the scalar monopole field we conclude
that $\Psi^{\text{S}}_{(3)}$ must be proportional to
\begin{displaymath}
A \epsilon_{j k l} K^j \dot{\mathcal{B}}^k{}_i  
\frac{x^l x^i}{\rho},
\end{displaymath}
where $A$ is a constant. Substituting this expression into the
differential equation for $\Psi^{\text{S}}_{(3)}$ gives that $A =
-1/7$, or that
\begin{equation}
\Psi^{\text{S}}_{(3)} = -\frac{1}{7} \epsilon_{jkl} K^j
\dot{\mathcal{B}}^k{}_i \frac{x^l x^i}{\rho}.
\end{equation}

Eq.~(\ref{OrderScalar}) gives that the third 
correction is of $O({\rho^2}/{{\mathcal{R}}^4})$.
This correction is not necessary for self-force calculations.
Differentiation of it would give terms proportional to the
coordinates, which would give zero after the limit to the location
of the particle is taken.

Finally, the singular scalar field of a dipole moving on a
geodesic is equal to
\begin{equation}
\Psi^{\text{S}} = \frac{K_i x^i}{\rho^3} - \frac{1}{7} \epsilon_{jkl}
\frac{K^j \dot{\mathcal{B}}^k{}_i x^l x^i}{\rho} +
O(\frac{\rho^2}{{\mathcal{R}}^4}).
\label{scalardipole}
\end{equation}
In this case, knowledge of the second correction is necessary 
in order to calculate the self-force. Differentiating the second
correction gives a term of order $O(\rho^0 / \mathcal{R}^3)$ which is
non-zero when the limit to the location of the particle is taken.

\subsection{Electromagnetic Potential}
\label{ElectromagneticPotential}

Assume that a particle carries an electric charge, an electric
dipole moment or a magnetic dipole moment. The source associated
with it is denoted by $J^a$. In this case, the self-force can be calculated
using the singular electromagnetic
potential, which can be expanded as
\begin{equation}
A_{\text{S}}^{a} = A_{\text{S} (0)}^{a} + A_{\text{S} (2)}^{a} +
A_{\text{S} (3)}^{a} + \ldots.
\label{expansionA}
\end{equation}
In the vacuum background the Ricci tensor is equal to zero.
That means that the electromagnetic potential obeys the Maxwell's
equations
\begin{equation}
\begin{split}
\nabla^{2}_{(0+2+3+\ldots)} A_{\text{S} (0+2+3+\ldots)}^{a}
        & = - 4\pi J^{a}.\\
\end{split}
\label{DelACharge}
\end{equation}

Regardless of the exact form of the source, the leading term
in the expansion (\ref{expansionA}) 
is the electromagnetic potential that would be generated by the
particle if it were stationary at the origin of a Cartesian
coordinate system. It obeys the equation
\begin{equation}
\nabla_{(0)}^{2}  A_{\text{S} (0)}^{a} = - 4 \pi J^{a}.
\label{DelACharge0}
\end{equation}
The first correction obeys the differential equation derived from
Eq.~(\ref{DelACharge}), namely
\begin{equation}
\nabla_{(0)}^{2} A_{\text{S} (2)}^{a} + \nabla_{(2)}^{2}
A_{\text{S} (0)}^{a} = 0. \label{DelACharge1}
\end{equation}
The $O(\rho^2 / \mathcal{R}^2)$-part of $\nabla^{2}$ acting
on the zeroth-order electromagnetic potential is the source term
for the first correction. From this equation it can also be
inferred that the order of the first correction to the electromagnetic
potential is
\begin{equation}
A_{\text{S} (2)}^{a} \sim [O( A_{\text{S}
(0)}^{a}) \times O (\frac{\rho^2}{\mathcal{R}^2}) ]. 
\end{equation}

According to Eq.~(\ref{DelACharge}), the second correction obeys the
differential equation
\begin{equation}
\nabla^{2}_{(0)} A_{\text{S} (3)}^{a} + \nabla^{2}_{(3)}
A_{\text{S} (0)}^{a}  = 0
\label{DelACharge2}
\end{equation}
meaning that the $O(\rho^3 / \mathcal{R}^3)$-part 
of $\nabla^2$ acting on the zeroth
order electromagnetic potential acts as the source term in the
differential equation for $A_{\text{S} (3)}^{a}$. As in the scalar
case, $A_{\text{S} (2)}^{a}$ does not show up in this equation,
but will show up in the equation for the third correction.
The equation for $A_{\text{S} (3)}^{a}$ indicates that 
\begin{equation}
A_{\text{S} (3)}^{a} \sim [ A_{\text{S} (0)}^{a} \times 
O(\frac{\rho^3}{\mathcal{R}^3}) ]
\end{equation}

The equation for the third correction as derived from Eq.~(\ref{DelACharge}) is
\begin{equation}
\nabla^2_{(0)}A_{\text{S} (4)}^{a} + \nabla^2_{(2)}A_{\text{S}
(2)}^{a} + \nabla^2_{(4)}A_{\text{S} (0)}^{a}
 = 0 \label{DelACharge3}
\end{equation}
which gives the order of the third correction
\begin{equation}
A_{\text{S} (4)}^{a} \sim [O(A_{\text{S}
(0)}^{a}) \times O(\frac{\rho^4}{\mathcal{R}^4}) ].
\label{OrderApotential3}
\end{equation}

\subsubsection{Monopole Potential} \label{SFECP}

If the particle is endowed with an electric charge $q$, 
Eq.~(\ref{DelACharge0}) gives that the
zeroth order potential is the Coulomb
electromagnetic potential 
\begin{equation}
A_{\text{S} (0)}^{a} = \Big ( \frac{q}{\rho}, \: 0,\: 0, \: 0 \Big ).
\end{equation}

Substituting the THZ components of $A_{\text{S} (2)}^{a}$
\begin{equation}
A^{a}_{\text{S} (2)} = ( A^{t}_{\text{S} (2)}, A^{x}_{\text{S} (2)},
A^{y}_{\text{S} (2)}, A^{z}_{\text{S} (2)} )
\end{equation}
into Eq.~(\ref{DelACharge1}) results in four
differential equations, one for each one of these components. Each
equation relates the sum of second derivatives of a
component to a sum of terms of the form:
\begin{displaymath}
q \mathcal{E}_{..} \frac{x^{.} x^{.} x^{.} x^{.}}{\rho^5} 
\end{displaymath}
for the $t$-component and 
\begin{displaymath}
q \mathcal{B}_{..} \frac{x^{.} x^{.}}{\rho^3} 
\end{displaymath}
for the spatial components,
where the dots again denote appropriate indices.

The form of the solution can be predicted based on the dimensionality and
the index structure as before. One must take into account the fact that
the $t$-component must have no free indices and that each spatial component 
must have one free index. Thus it can be derived that the solution should
be proportional to
\begin{displaymath}
q \mathcal{E}_{ij} \frac{x^i x^j}{\rho} 
\end{displaymath}
for the $t$-component and to
\begin{displaymath}
q \epsilon^{p}{}_{ij} \mathcal{B}^{i}{}_k \frac{x^k x^j}{\rho} 
\end{displaymath}
for the $p$-spatial component,
each term multiplied by an appropriate constant. Substituting
these expressions into the four differential equations gives 
simple algebraic equations for these constants, which can be
easily solved to give the components of the first correction 
\begin{equation}
\begin{split}
A_{\text{S} (2)}^{t} &= -\frac{1}{2} q \mathcal{E}_{ij} \frac{x^i x^j}{\rho},\\
A_{\text{S} (2)}^{p} &= \frac{1}{2} q
\epsilon^{p}{}_{ij} \mathcal{B}^{i}{}_ k \frac{x^j x^k}{\rho}. \\
\end{split}
\end{equation}

The second correction to the electromagnetic potential
comes from the part of the $\nabla^2$
that is of $O(\frac{\rho^3}{\mathcal{R}^3})$ acting on the
zeroth-order electromagnetic potential. Assuming that the THZ
components of $A_{\text{S} (3)}^a$ are
\begin{equation}
A_{\text{S} (3)}^a = ( A_{\text{S} (3)}^t, A_{\text{S} (3)}^x,
    A_{\text{S} (3)}^y ,A_{\text{S} (3)}^z)
\end{equation}
and substituting into Eq.~(\ref{DelACharge2}) gives a differential
equation for each component of $A_{\text{S} (3)}^{a}$. Each
equation relates a sum of second derivatives of each component to terms
of the form
\begin{displaymath}
q {\mathcal{E}}_{...} \frac{x^. x^. x^. x^. x^.}{\rho^5} 
\end{displaymath}
for the $t$-component, and
\begin{displaymath}
q \dot{\mathcal{E}}_{..} \frac{x^. x^. x^.}{\rho^3} 
\text{ and }
q \mathcal{B}_{...} \frac{x^. x^. x^.}{\rho^3}
\end{displaymath}
for the spatial components.
The $t$-component in this case is expected to be proportional to
\begin{displaymath}
q \mathcal{E}_{ijk} \frac{x^i x^j x^k}{\rho}
\end{displaymath}
and the $p$-spatial component is expected to be a sum of the terms 
\begin{displaymath}
\begin{split}
& q \dot{\mathcal{E}}^p{}_i x^i \rho , \:
q \dot{\mathcal{E}}_{ij} \frac{x^p x^i x^j}{\rho} \\
& q \epsilon^p{}_{ij} \mathcal{B}^{ij}{}_k x^k \rho , \:
q \epsilon^p{}_{ij} \mathcal{B}^i{}_{kl} \frac{x^j x^k x^l}{\rho} ,\:
q \epsilon_{ijk} \mathcal{B}^{ik}{}_l \frac{x^p x^j x^l}{\rho} \\
\end{split}
\end{displaymath}
each term multiplied by an appropriate constant. Substituting
these expressions into the four differential equations results in
algebraic equations for the constants. Finally, the second
correction to the electromagnetic potential is
\begin{equation}
\begin{split}
A_{\text{S} (3)}^t = &-\frac{1}{6} q 
                       \mathcal{E}_{ijk} \frac{x^i x^j x^k}{\rho}\\
A_{\text{S} (3)}^p = &-\frac{1}{18} q \dot{\mathcal{E}}^p{}_i x^i
\rho
+ \frac{7}{18} q \dot{\mathcal{E}}_{ij} \frac{x^p x^i x^j}{\rho} \\
& +\frac{2}{9} q \epsilon^p{}_{ij} \mathcal{B}^i{}_{kl}
 \frac{x^j x^k x^l}{\rho} .\\
\end{split}
\end{equation}

According to Eq.~(\ref{OrderApotential3}) the order of the
third correction to the electromagnetic potential of a charge is
\begin{equation}
A_{\text{S (4)}}^a \sim O(\frac{\rho^3}{\mathcal{R}^4}).
\end{equation}
and it is not necessary for self-force calculations.

Finally, the components of the singular electromagnetic potential of a charge
$q$ that is moving on a geodesic are equal to
\begin{equation}
\begin{split}
A_{\text{S}}^{t} =&\frac{q}{\rho} - \frac{1}{2} q \mathcal{E}_{ij} 
                   \frac{x^i x^j}{\rho}
                   -\frac{1}{6} q
                    \mathcal{E}_{ijk} \frac{x^i x^j x^k}{\rho}
                    + O(\frac{\rho^3}{\mathcal{R}^4}), \\
A_{\text{S}}^{p} = &\frac{1}{2} \epsilon^{p}{}_{ij} q
              \mathcal{B}^{i}{}_k \frac{x^j x^k}{\rho}  \\
              & -\frac{1}{18} q \dot{\mathcal{E}}^p{}_i x^i \rho
             + \frac{7}{18} q \dot{\mathcal{E}}_{ij} 
             \frac{x^p x^i x^j}{\rho} \\
             & +\frac{2}{9} q \epsilon^p{}_{ij} \mathcal{B}^i_{kl}
             \frac{x^j x^k x^l}{\rho} + O(\frac{\rho^3}{\mathcal{R}^4}).\\
\end{split}
\end{equation}
In this case, the second correction is not necessary 
for the calculation of the self-force. After differentiation, that correction 
will give terms proportional to the coordinates which, after taking
the limit to the location of the particle, will give zero.

\subsubsection{Electric Dipole Potential}
\label{SFED}

Assume that a particle carrying an electric dipole moment is moving on a
geodesic $\Gamma$.
The dipole moment is assumed to point at some random direction and its THZ
components are $q^{a} = (0, q^x, q^y, q^z)$.

According to Eq.~(\ref{DelACharge0})
the leading term in the expansion of the singular 
electromagnetic potential is the 
electromagnetic potential generated by an electric
dipole that is stationary at the origin of the Cartesian coordinate system
\begin{equation}
A^{a}_{\text{S} (0)} = \bigg ( \frac{q_i x^i}{\rho^{3}}, \: 0, \: 0, \: 0
\bigg ).
\end{equation}

The THZ components of the first correction are
\begin{equation}
A^{a}_{\text{S} (2)} = ( A^{t}_{\text{S} (2)}, A^{x}_{\text{S}
(2)}, A^{y}_{\text{S} (2)}, A^{z}_{\text{S} (2)}).
\end{equation}
If substituted into Eq.~(\ref{DelACharge1}), the differential
equation for $A_{\text{S} (2)}^{a}$ is broken into a set of four
second-order differential equations for these components. Each
differential equation relates the sum of second derivatives of a
component to a sum of terms of the form
\begin{displaymath}
q_. \mathcal{E}_{..} \frac{x^. x^. x^. x^. x^.}{\rho^7} 
\end{displaymath}
for the $t$-component, 
\begin{displaymath}
q_. \mathcal{B}_{..} \frac{x^. x^. x^. x^. x^.}{\rho^7} 
\end{displaymath}
for the spatial components.

In this case we can deduce that the solution should be equal to a
sum of the terms
\begin{displaymath}
q_i \mathcal{E}_{jk} \frac{x^i x^j x^k}{\rho^3}, \: q^i
\mathcal{E}_{ij} \frac{x^j}{\rho}
\end{displaymath}
for the $t$-component and a sum of the terms
\begin{displaymath}
\begin{split}
&\epsilon^{p}{}_{ij} q^{i} \mathcal{B}^{j}{}_l \frac{x^l}{\rho}, \:
\epsilon^{p}{}_{ij} q^{i} \mathcal{B}_{kl} \frac{x^j x^k x^l}{\rho^3},\:
\epsilon^{p}{}_{ij} q^{k} \mathcal{B}^{i}{}_{k} \frac{x^j}{\rho},\\
&\epsilon^{p}{}_{ij} q_{k} \mathcal{B}^{i}{}_{l} \frac{x^j x^k x^l}{\rho^3}, \:
\epsilon_{ijk} q^{i} \mathcal{B}^{pj} \frac{x^k}{\rho}, \:
\epsilon_{ijk} q^{i} \mathcal{B}^{j}{}_l \frac{x^p x^k x^l}{\rho^3}, \\
\end{split}
\end{displaymath}
for the $p$-spatial component, each term multiplied by an
appropriate constant so that the differential equations are satisfied.
Additionally, using Eq.~(\ref{EBRel2}) and
(\ref{EBRel4}), the last two terms that are expected to show up in
the solution for the $p$-component can be eliminated
 in favor of the remaining four.
Substituting the sums of the remaining terms
into the four differential equations
gives simple algebraic equations for the multiplicative constants. The final
expressions for the components of $A^{a}_{\text{S} (2)}$ are
\begin{equation}
\begin{split}
A^{t}_{\text{S} (2)} = &-\frac{1}{2} q_i \mathcal{E}_{jk} \frac{x^i x^j x^k}
{\rho^3}, \\
A^{p}_{\text{S} (2)} = &\frac{1}{2} \epsilon^{p}{}_{ij} q^{i} \mathcal{B}^{j}{}_{l}
\frac{x^l}{\rho} + 
\frac{1}{2}\epsilon^{p}{}_{ij} q^{k} \mathcal{B}^{i}{}_{k} \frac{x^j}{\rho} \\
   &+ \frac{1}{2} \epsilon^{p}{}_{ij} q_{k} \mathcal{B}^{i}{}_{l}
\frac{x^j x^k x^l}{\rho^3}.
\end{split}
\end{equation}

The second correction to the singular electromagnetic potential obeys
Eq.~(\ref{DelACharge2}). By substituting its THZ components
\begin{equation}
A_{\text{S} (3)}^a = ( A_{\text{S} (3)}^t, A_{\text{S} (3)}^x,
    A_{\text{S} (3)}^y ,A_{\text{S} (3)}^z)
\end{equation}
into (\ref{DelACharge2}) we get a set of four second-order
differential equations for each one of those components. Each
equation relates the second derivatives of one component to terms
of the form
\begin{displaymath}
 q_. \dot{\mathcal{B}}_{..} \frac{x^. x^. x^. x^. x^. x^.}{\rho^7},
 q_. \mathcal{E}_{...} \frac{x^. x^. x^. x^. x^. x^.}{\rho^7}
\end{displaymath}
for the $t$-component, and to terms of the form
\begin{displaymath}
q_. \dot{\mathcal{E}}_{..} \frac{x^. x^. x^. x^.}{\rho^5} ,
q_. \mathcal{B}_{...} \frac{x^. x^. x^. x^.}{\rho^5} ,
\end{displaymath}
for each spatial component. 

In this case the solution must be a sum of the terms
\begin{displaymath}
\begin{split}
& \epsilon_{ijk} q^i \dot{\mathcal{B}}^j{}_{l} \frac{x^k x^l}{\rho}, \\
 & q_i \mathcal{E}_{jkl} \frac{x^i x^j x^k x^l}{\rho^3}, \:
 q^i \mathcal{E}_{ijk} \frac{x^j x^k}{\rho} , \\
\end{split}
\end{displaymath}
for the $t$-component, and a sum of terms
\begin{displaymath}
\begin{split}
& q^p \dot{\mathcal{E}}_{ij} \frac{x^i x^j}{\rho} ,\:
  q_j \dot{\mathcal{E}}^p{}_i \frac{x^i x^j}{\rho},\:
  q^i \dot{\mathcal{E}}^p{}_i \rho \\
& q^i \dot{\mathcal{E}}_{ij} \frac{x^p x^j}{\rho},\:
  q^i \dot{\mathcal{E}}_{jk} \frac{x^p x_i x^j x^k}{\rho^3} \\
& \epsilon^p{}_{ij} q^i \mathcal{B}^j{}_{kl} \frac{x^k x^l}{\rho},\:
  \epsilon^p{}_{ij} q^i \mathcal{B}_{klm} \frac{x^j x^k x^l x^m}{\rho^3}, \:
  \epsilon^p{}_{ij} q^k \mathcal{B}^i{}_{kl} \frac{x^j x^l}{\rho} , \\ 
&  \epsilon^p{}_{ij} q^k \mathcal{B}^i{}_{lm} \frac{x^j x_k x^l x^m}{\rho^3},\:
 \epsilon_{ijk} q^i \mathcal{B}^j{}_{lm} \frac{x^p x^k x^l x^m}{\rho^3} \\
\end{split}
\end{displaymath}
for the $p$-spatial component, each term multiplied by an
appropriate constant so that the differential equations are
satisfied. We substitute these expressions into the
differential equations and solve the resulting systems of algebraic
equations for those constants. Thus, the second
correction to the electromagnetic potential is
\begin{equation}
\begin{split}
A^t_{\text{S (3)}} =& -\frac{1}{7}  \epsilon_{ijk} q^i \dot{\mathcal{B}}^j{}_l 
  \frac{x^k x^l}{\rho} - \frac{1}{6} q_i \mathcal{E}_{jkl} 
  \frac{x^i x^j x^k x^l}{\rho^3} \\
A^p_{\text{S (3)}} = &-\frac{1}{18}  q^p
\dot{\mathcal{E}}_{ij} \frac{x^i x^j}{\rho} -\frac{5}{18} q_j
\dot{\mathcal{E}}^p{}_i \frac{x^i x^j}{\rho} 
    + \frac{2}{9} q^i \dot{\mathcal{E}}^p{}_i \rho \\
    &+ \frac{1}{18}
   q^i \dot{\mathcal{E}}_{ij} \frac{x^p x^j}{\rho} 
   +\frac{7}{18} q^i \dot{\mathcal{E}}_{jk} \frac{x^p x_i x^j x^k}{\rho^3} \\
   & +\frac{2}{9}\epsilon^p{}_{ij} q^k \mathcal{B}^i{}_{lm} \frac{x^j
   x_k x^l x^m}{\rho^3} + \frac{2}{9}  \epsilon_{ijk} q^i
   \mathcal{B}^{pj}{}_l \frac{x^k x^l}{\rho} . \\
\end{split}
\end{equation}

According to Eq.~(\ref{OrderApotential3}) the third correction is 
\begin{equation}
A^a_{\text{S (4)}} \sim O(\frac{\rho^2}{\mathcal{R}^4})
\end{equation}
which is clearly not necessary for self-force calculations, because taking
the first derivative and the limit to the location of the particle gives
zero.

Finally, the singular electromagnetic potential for an electric dipole moving
on a geodesic is equal to
\begin{equation}
\begin{split}
A^t_{\text{S}} = & q_i \frac{x^i}{\rho^3} -\frac{1}{2} q_i
\mathcal{E}_{jk} \frac{x^i x^j x^k}{\rho^3}   \\
  &-\frac{1}{7} \epsilon_{ijk}
q^i \dot{\mathcal{B}}^j{}_l \frac{x^k x^l}{\rho} - \frac{1}{6} 
q_i \mathcal{E}_{jkl} \frac{x^i x^j x^k x^l}{\rho^3} 
 + O(\frac{\rho^2}{\mathcal{R}^4}) \\
A^p_{\text{S}} = &\frac{1}{2} \epsilon^{p}{}_{ij}
q^{i} \mathcal{B}^{j}{}_{l} \frac{x^l}{\rho} 
   + \frac{1}{2} \epsilon^{p}{}_{ij} q^{k}
\mathcal{B}^{i}{}_{k} \frac{x^j}{\rho} \\
   &+ \frac{1}{2} \epsilon^{p}{}_{ij} q_{k} \mathcal{B}^{i}{}_{l} 
   \frac{x^j x^k x^l}{\rho^3} \\
   &-\frac{1}{18} q^p \dot{\mathcal{E}}_{ij} \frac{x^i x^j}{\rho} 
-\frac{5}{18}  q_j \dot{\mathcal{E}}^p{}_i \frac{x^i x^j}{\rho} 
 + \frac{2}{9} q^i \dot{\mathcal{E}}^p{}_i \rho \\ 
 &+ \frac{1}{18} q^i \dot{\mathcal{E}}_{ij} \frac{x^p x^j}{\rho} 
    +\frac{7}{18}  q^i \dot{\mathcal{E}}_{jk} 
    \frac{x^p x_i x^j x^k}{\rho^3} \\
   & + \frac{2}{9} \epsilon^p{}_{ij} q^k \mathcal{B}^i{}_{lm} 
   \frac{ x^j x_k x^l x^m}{\rho^3} 
   + \frac{2}{9}  \epsilon_{ijk} q^i
   \mathcal{B}^{pj}{}_l \frac{x^k x^l}{\rho}  \\
& + O(\frac{\rho^2}{\mathcal{R}^4}). \\
\end{split}
\end{equation}
In this case the second correction is necessary
for the self-force calculation. That correction is proportional to the 
first power of the coordinates, differentiation of which gives a constant
term
that does not vanish when the limit to the location of the particle is 
taken.

\subsubsection{Magnetic Dipole Potential} \label{SFMD}

Assume that a particle with a given magnetization is moving on
the geodesic $\Gamma$. The magnetization $m^{a}$ is assumed to
point at some random direction and its THZ components are $m^{a} =
(0, m^x, m^y, m^z)$.

Eq.~(\ref{DelACharge0}) gives that
the leading term in the expansion of the electromagnetic potential
is the potential generated by a magnetic
dipole that is stationary at the origin of a Cartesian coordinate system. 
Its THZ components are
\begin{equation}
A^{a}_{\text{S} (0)} = ( 0, \epsilon^x{}_{ij} m^i \frac{x^j}{\rho^3},
\epsilon^y{}_{ij} m^i \frac{x^j}{\rho^3},\epsilon^z{}_{ij}
m^i \frac{x^j}{\rho^3}).
\end{equation}

The first correction $A_{\text{S (2)}}^{a}$ has THZ components:
\begin{equation}
A^{a}_{\text{S} (2)} = ( A^{t}_{\text{S} (2)}, A^{x}_{\text{S}
(2)}, A^{y}_{\text{S} (2)}, A^{z}_{\text{S} (2)} ).
\end{equation}
When those componenets are
substituted into Eq.~(\ref{DelACharge1}) the result is a set
of four second-order differential equations, one for each one of
those four components. Each equation relates a sum of the second
derivatives of a component to a source term that consists of
terms of the form
\begin{displaymath}
 m_{.} \mathcal{B}_{..} \frac{x^. x^. x^.}{\rho^5} 
\end{displaymath}
for the $t$-component, and terms of the form
\begin{displaymath}
m_{.} \mathcal{E}_{..} \frac{x^. x^. x^. x^. x^.}{\rho^7} 
\end{displaymath}
for the spatial components,
where the dots denote appropriate indices.

The form of the solution can be predicted as previously. In this case, the
$t$-component is expected to be a sum of the terms
\begin{displaymath}
m_{i} \mathcal{B}^{i}{}_{j} \frac{x^{j}}{\rho} ,
\quad \frac{m_{i} \mathcal{B}_{jk} x^{i} x^{j} x^{k}}{\rho^3}
\end{displaymath}
and the $p$-spatial component a sum of the terms
\begin{displaymath}
\begin{split}
&\epsilon^{p}{}_{ij} m^{i} \mathcal{E}^{j}{}_{k} \frac{x^{k}}
{\rho}, \: \epsilon^{p}{}_{ij} m^{i} \mathcal{E}_{kl} \frac{x^{j} x^{k}
x^{l}}{\rho^3}, \:
\epsilon^{p}{}_{ij} m^{k} \mathcal{E}^{i}{}_{k} \frac{x^{j}}
{\rho} , \\
&\epsilon^{p}{}_{ij} m^{k} \mathcal{E}^{i}{}_{l} \frac{x^{j} x_{k} x^{l}}
{\rho^3}, \:
\epsilon_{ijk} m^{i} \mathcal{E}^{pj} \frac{x^{k}}{\rho}, \:
\epsilon_{ijk} m^{i} \mathcal{E}^{j}{}_{l} \frac{x^{k} x^{l} x^{p}}{\rho^3} \\
\end{split}
\end{displaymath}
with appropriate constants multiplying
each term so that the differential equations are satisfied. Using 
Eq.~(\ref{EBRel2}) and (\ref{EBRel4}), the first and last terms
expected to show up in the sum for the $p$-component
can be eliminated, since they can be expressed as linear
combinations of the remaining four terms. Substituting these
expressions into the differential equations gives simple systems of 
algebraic equations for those constants.
The result is that the first correction to the electromagnetic potential 
has components:
\begin{equation}
\begin{split}
A^{t}_{\text{S} (2)}= & \frac{1}{6} m_{i}
\mathcal{B}_{jk} \frac{x^i x^j x^k}{\rho^3} - \frac{2}{3} 
m^{i} \mathcal{B}_{ij} \frac{x^j}{\rho}  \\
A^{p}_{\text{S} (2)} = & \epsilon^{p}{}_{ij}  m^{i}
\mathcal{E}_{kl} \frac{x^j x^k x^l}{\rho^3} - \frac{1}{2} 
\epsilon^{p}{}_{ij} m_{k}
\mathcal{E}^{i}{}_{l} \frac{x^{j} x^{k} x^{l}}{\rho^3} \\
        &- \frac{1}{2} \epsilon_{ijk} m^{i} \mathcal{E}^{pj}
        \frac{x^k}{\rho}  . \\
\end{split}
\end{equation}

The second correction to the magnetic dipole potential can
be calculated using Eq.~(\ref{DelACharge2}). Substituting the THZ
components of that correction:
\begin{equation}
A^{a}_{\text{S} (3)} = ( A^{t}_{\text{S} (3)}, A^{x}_{\text{S}
(3)}, A^{y}_{\text{S} (3)}, A^{z}_{\text{S} (3)} )
\end{equation}
to Eq.~(\ref{DelACharge2}) we get four differential equations, one
for each component, which relate second derivatives of those
components to terms of the form
\begin{displaymath}
m_. \dot{\mathcal{E}}_{..} \frac{x^. x^. x^. x^.}{\rho^5} ,
m_. \mathcal{B}_{...} \frac{x^. x^. x^. x^.}{\rho^5}
\end{displaymath}
for the $t$-component and terms of the form
\begin{displaymath}
m_. \dot{\mathcal{B}}_{..} \frac{x^. x^. x^. x^. x^.x^.}{\rho^7} , 
m_. \mathcal{E}_{...} \frac{x^. x^. x^. x^. x^. x^.}{\rho^7}
\end{displaymath}
for the spatial components. 
The terms expected to show up in the solution can be constructed
as before. We expect a sum of the terms
\begin{displaymath}
\begin{split} 
& \epsilon_{ijk} m^i \dot{\mathcal{E}}^j{}_l \frac{x^k x^l}{\rho}, \\
& m^i \mathcal{B}_{ijk} \frac{x^j x^k}{\rho} , 
m_l \mathcal{B}_{ijk} \frac{x^i x^j x^k x^l}{\rho^3}
\end{split}
\end{displaymath}
for the $t$-component and a sum of the terms
\begin{displaymath}
\begin{split}
& m^p \dot{\mathcal{B}}_{ij} \frac{x^i x^j}{\rho} , \: m^i
\dot{\mathcal{B}}^p{}_i \rho,\:
 m_j \dot{\mathcal{B}}^p{}_i \frac{x^i x_j}{\rho} ,\\
& m_i \dot{\mathcal{B}}^i{}_j \frac{x^p x^j}{\rho}, \:
 \frac{1}{\rho^3} m_i \dot{\mathcal{B}}_{jk} x^p x^i x^j x^k \\
& \epsilon^p{}_{ij} m^i \mathcal{E}^j{}_{kl} \frac{x^k
x^l}{\rho} , \: \epsilon^p{}_{ij} m^k \mathcal{E}^i{}_{kl}
\frac{x^j x^l}{\rho}, \\
& \epsilon_{ijk} m^i \mathcal{E}^{pj}{}_l \frac{x^k x^l}{\rho} ,\:
 \epsilon_{ijk} m^i \mathcal{E}^j{}_{ln} \frac{x^p x^k
x^l x^n}{\rho^3}, \\
 & \epsilon^p{}_{ij} m^i \mathcal{E}_{kln} \frac{x^j x^k
x^l x^n}{\rho^3}, \:
  \epsilon^p{}_{ij} m_n
\mathcal{E}^i{}_{kl} \frac{x^j x^k x^l x^n}{\rho^3}
\end{split}
\end{displaymath}
for the $p$-spatial component, each term multiplied by an
appropriate constant. Substituting those sums into the
differential equations, we get systems of algebraic equations for
those constants. Solving those we get that the
second correction to the electromagnetic potential is 
\begin{equation}
\begin{split}
A^{t}_{\text{S} (3)} &= \frac{1}{7}  \epsilon_{ijk}
m^i \dot{\mathcal{E}}^j{}_l \frac{x^k x^l}{\rho} \\
 & -\frac{1}{3} m^i \mathcal{B}_{ijk} \frac{x^j x^k}{\rho} +
 \frac{1}{9} m_l \mathcal{B}_{ijk} \frac{x^i x^j x^k x^l}
{\rho^3} \\
A^{p}_{\text{S} (3)} &= \frac{37}{252} m^p \dot{\mathcal{B}}_{ij}
\frac{x^i x^j}{\rho} + \frac{25}{126} m^i
\dot{\mathcal{B}}^p{}_i
\rho + \frac{5}{63} m^i \dot{\mathcal{B}}_{ij} \frac{x^p x^j}
{\rho}\\
  & - \frac{4}{63} m^i \dot{\mathcal{B}}^p{}_j \frac{x_i
  x^j}{\rho}  
   -\frac{1}{36} m_i \dot{\mathcal{B}}_{jk} \frac{x^p x^i x^j x^k}
  {\rho^3} \\
  & +\frac{1}{6} \epsilon^p{}_{ij} m^i \mathcal{E}_{kln} \frac{x^j x^k
  x^l x^n}{\rho^3}  -\frac{1}{6} \epsilon^p{}_{ij} m^k
  \mathcal{E}^i{}_{kl} \frac{x^j x^l}{\rho} \\
  & -\frac{1}{6} \epsilon_{ijk} m^i \mathcal{E}^j{}_{ln} \frac{x^p x^k
  x^l x^n}{\rho^3}. \\
\end{split}
\end{equation}

The order of the third correction can be predicted as usual,
by Eq.~(\ref{DelACharge3}) and it is
\begin{equation}
A^{p}_{\text{S} (4)} \sim O(\frac{\rho^2}{\mathcal{R}^4}).
\end{equation}
These correction terms are not needed for self-force calculations.
As in the case of the electric dipole electromagnetic potential, 
differentiating these terms would result in terms that are proportional
to the cooridinates and taking the limit to the location of the particle
gives zero.

Finally, the singular electromagnetic potential generated by 
a magnetic dipole moving on a geodesic $\Gamma$ has components equal to
\begin{equation}
\begin{split}
A^{t}_{\text{S}} &=  \frac{1}{6}  m_{i}
\mathcal{B}_{jk} \frac{x^{i} x^{j} x^{k}}{\rho^3} -
 \frac{2}{3} m^{i} \mathcal{B}_{ij} \frac{x^j}{\rho} \\
 & +\frac{1}{7}  \epsilon_{ijk}
 m^i \dot{\mathcal{E}}^j{}_l \frac{x^k x^l}{\rho} \\
 & -\frac{1}{3} m^i \mathcal{B}_{ijk} \frac{x^j x^k}{\rho} +
 \frac{1}{9} m_l \mathcal{B}_{ijk} \frac{x^i x^j x^k x^l}{\rho^3} 
 +O(\frac{\rho^2}{\mathcal{R}^4}) \\
A^{p}_{\text{S}} &= \epsilon^{p}{}_{ij} \frac{m^i x^j}{\rho^3} + 
\epsilon^{p}{}_{ij} \frac{m^{i} \mathcal{E}_{kl} x^{j} x ^{k}
x^{l}}{\rho^3}  \\& - \frac{1}{2} \epsilon^{p}{}_{ij} \frac{m_{k}
\mathcal{E}^{i}{}_{l} x^{j} x^{k} x^{l}}{\rho^3}
 -\frac{1}{2} \epsilon_{ijk} m^{i} \mathcal{E}^{pj} \frac{x^k} 
 {\rho}\\
 &+\frac{37}{252} m^p \dot{\mathcal{B}}_{ij}
 \frac{x^i x^j}{\rho} + \frac{25}{126} m^i
\dot{\mathcal{B}}^p{}_i
\rho \\
  & - \frac{4}{63} m^i \dot{\mathcal{B}}^p{}_j \frac{x_i
  x^j}{\rho} + \frac{5}{63} m^i \dot{\mathcal{B}}_{ij} \frac{x^p x^j}{\rho}\\
  & -\frac{1}{36} m_i \dot{\mathcal{B}}_{jk} \frac{x^p x^i x^j x^k}
  {\rho^3} \\
  & +\frac{1}{6} \epsilon^p{}_{ij} m^i \mathcal{E}_{kln} \frac{x^j x^k
  x^l x^n}{\rho^3}  -\frac{1}{6} \epsilon^p{}_{ij} m^k
  \mathcal{E}^i{}_{kl} \frac{x^j x^l}{\rho} \\
  & -\frac{1}{6} \epsilon_{ijk} m^i \mathcal{E}^j{}_{ln} \frac{x^p x^k
  x^l x^n}{\rho^3} + O(\frac{\rho^2}{\mathcal{R}^4}). \\
\end{split}
\end{equation}
It is clear that in the case of the magnetic dipole ,
it is necessary to know the second correction to the singular
electromagnetic potential in order to be able to calculate the
self-force.

\subsection{Gravitational Field}
\label{GravitationalField}

We proceed by looking into the more interesting (and computationally
more complicated) case of a particle moving on a
background geodesic $\Gamma$ and producing a gravitational field.
The singular gravitational field can be expanded as
\begin{equation}
h_{\text{S}}{}_{ab} = h^{(0)}_{\text{S}}{}_{ab} +
h^{(2)}_{\text{S}}{}_{ab} + h^{(3)}_{\text{S}}{}_{ab} + \hdots .
\end{equation}
For the calculation of the gravitational singular field it is more
convenient to use the trace-reversed version of $h_{\text{S}}{}_{ab}$
which is defined as
\begin{equation}
\bar{h}_{\text{S}}{}_{ab} = h_{\text{S} \: ab} - \frac{1}{2} g_{ab}
h_{\text{S}}{}^c{}_c .
\end{equation}
We work in the harmonic gauge 
\begin{equation}
\nabla^a \bar{h}_{\text{S}}{}_{ab} = 0.
\label{HarmonicGauge}
\end{equation}
The singular field obeys the linearized Einstein equations which,
in the harmonic gauge, become
\begin{equation}
\begin{split}
&\nabla^2_{(0+2+3+\hdots)} \bar{h}^{(0+2+3+\hdots)}_{\text{S}}{}_{ab} \\
&+ 2
R_{(0+2+3+\hdots)}{}_{a}{}^c{}_b{}^d \bar{h}^{(0+2+3+\hdots)}_{\text{S}}{}_{cd}
=  - 16 \pi T_{ab}. \\
\end{split}
\label{LinEinstein}
\end{equation}
In Eq.~(\ref{LinEinstein}) $T_{ab}$
is the stress-energy tensor associated with the particle.

Regardless of the exact form of the stress-energy tensor, the
leading term in the expansion of the gravitational field is
the field that the particle would produce if it were stationary at
the origin of a Cartesian coordinate system and it obeys the
equation
\begin{equation}
\nabla^2_{(0)} \bar{h}^{(0)}_{\text{S}}{}_{ab} + 2
R_{(0)}{}_{a}{}^c{}_b{}^d \bar{h}^{(0)}_{\text{S}}{}_{cd} = - 16 \pi T_{ab} .
\label{LinEinstein0}
\end{equation}

The first correction obeys the differential
equation derived from (\ref{LinEinstein}) by keeping only terms that
come from the $O(\rho^2 / \mathcal{R}^2)$ part of the metric, namely
\begin{equation}
\begin{split}
& \nabla^2_{(0)} \bar{h}^{(2)}_{\text{S}}{}_{ab} + 2
R_{(0)}{}_{a}{}^c{}_b{}^d \bar{h}^{(2)}_{\text{S}}{}_{cd} = \\
& -\nabla^2_{(2)} \bar{h}^{(0)}_{\text{S}}{}_{ab} - 2
R_{(2)}{}_{a}{}^c{}_b{}^d \bar{h}^{(0)}_{\text{S}}{}_{cd} .\\
\end{split}
\label{LinEinstein1}
\end{equation}
The $\nabla^2$ and Riemann tensors of $O(\rho^2 / \mathcal{R}^2)$ acting on the
zeroth order gravitational field act as the source terms in the
differential equation of the first correction.
Eq.~(\ref{LinEinstein1}) also implies that the first 
correction is
\begin{equation}
h^{(2)}_{\text{S}}{}_{ab} \sim \big [ O(h^{(0)}_{\text{S}}{}_{ab}) \times
    O(\frac{\rho^2}{\mathcal{R}^2}) \big ].
\end{equation}

Similarly, the second correction obeys the
differential equation derived from Eq.~(\ref{LinEinstein}) by 
keeping only terms 
that come from the $O(\rho^3 / \mathcal{R}^3)$ part of the metric, namely
\begin{equation}
\begin{split}
& \nabla^2_{(0)} \bar{h}^{(3)}_{\text{S}}{}_{ab} + 2
R_{(0)}{}_{a}{}^c{}_b{}^d \bar{h}^{(3)}_{\text{S}}{}_{cd} = \\
& -\nabla^2_{(3)} \bar{h}^{(0)}_{\text{S}}{}_{ab} - 2
R_{(3)}{}_{a}{}^c{}_b{}^d \bar{h}^{(0)}_{\text{S}}{}_{cd}. \\
\end{split}
\label{LinEinstein2}
\end{equation}
Eq.~(\ref{LinEinstein2}) also implies that the second
correction is
\begin{equation}
h^{(3)}_{\text{S}}{}_{ab} \sim \big [ O(h^{(0)}_{\text{S}}{}_{ab}) \times
    O(\frac{\rho^3}{\mathcal{R}^3}) \big ].
\end{equation}

The order of the third correction to the gravitational field can
also be predicted, based on the differential equation that could
potentially yield that correction. Specifically,
Eq.~(\ref{LinEinstein}) gives
\begin{equation}
\begin{split}
& \nabla^2_{(0)} \bar{h}^{(4)}_{\text{S}}{}_{ab} + 2
R_{(0)}{}_{a}{}^c{}_b{}^d \bar{h}^{(4)}_{\text{S}}{}_{cd} = \\
& -\nabla^2_{(4)} \bar{h}^{(0)}_{\text{S}}{}_{ab} - 2
R_{(4)}{}_{a}{}^c{}_b{}^d \bar{h}^{(0)}_{\text{S}}{}_{cd} \\
& -\nabla^2_{(2)} \bar{h}^{(2)}_{\text{S}}{}_{ab} - 2
R_{(2)}{}_{a}{}^c{}_b{}^d \bar{h}^{(2)}_{\text{S}}{}_{cd} \\
\end{split}
\label{LinEinstein3}
\end{equation}
which implies that the third correction is of order
\begin{equation}
h^{(4)}_{\text{S}}{}_{ab} \sim \big [ O(h^{(0)}_{\text{S}}{}_{ab} ) \times
    O(\frac{\rho^4}{\mathcal{R}^4}) \big ].
\end{equation}

\subsubsection{Massive Particle}

Assume that the particle that generates the gravitational field is
a massive point particle of mass $m$. 
The calculation for the first correction to the gravitational singular field 
for this case was first shown in \cite{Det2005}, where a different
coordinate system was used. 
Here we perform the calculation of the singular field in the THZ coordinate 
system, including up to the second correction to it. 

The leading term in the expansion of the singular gravitational field is
the gravitational field generated by a massive particle that is stationary
at the origin of the Cartesian coordinate system. We consider only the terms
that are of first-order in the mass $m$. Thus
\begin{equation}
h^{(0)}_{\text{S}}{}_{ab} = 2 \frac{m}{\rho}
\left(\begin{array}{cccc}
1 & 0 & 0 & 0 \\
0 & 1 & 0 & 0 \\
0 & 0 & 1 & 0 \\
0 & 0 & 0 & 1
\end{array} \right).
\end{equation}

The first correction to this gravitational field obeys Eq.~(\ref{LinEinstein1}) 
and must be a symmetric tensor. Substituting its components into
Eq.~(\ref{LinEinstein1}) results in ten second-order differential equations.
There is one independent set of four differential equations for the four
diagonal components, each equation containing all four diagonal components.
There is also one differential equation for each of the $t - i$ components
and one differential equation for each of the spatial $i-j$ components
with $i \neq j$. In each equation a sum of second derivatives of components
is related to a sum of terms of the form
\begin{displaymath}
m \mathcal{E}_{..} \frac{x^. x^. x^. x^. x^. x^.}{\rho^7},
\end{displaymath}
for the $t-t$ and $p-q$ components and a sum of terms
\begin{displaymath}
m \mathcal{B}_{..} \frac{x^. x^. x^. x^. x^. x^.}{\rho^7},
\end{displaymath}
for the $t-p$ components.

Solving the differential equations in this case is more complicated
than in the previous cases, mainly because of the fact that four of them
involve all diagonal components rather than only one of them. However, the 
process that was described at the beginning of Sec.~\ref{Calc} for
predicting the form of the solution is applicable in this case as well. 
It is clear that the $t-t$ component must have no free indices, each of
the $t-p$ components must have one free index and each purely spatial
component must have two free indices. We conclude that the $t-t$
component of the first correction to the singular field must be
proportional to
\begin{displaymath}
m \mathcal{E}_{ij} \frac{x^i x^j}{\rho},
\end{displaymath}
each $t-p$ component must be proportional to
\begin{displaymath}
m \epsilon_{p i j} \mathcal{B}^i{}_k \frac{x^j x^k}{\rho}
\end{displaymath}
and each purely spatial $p-q$ component must be a sum of the terms
\begin{displaymath}
\begin{split}
&m \mathcal{E}_{pq} \rho, \: m \mathcal{E}_{pi} \frac{x_q x^i}{\rho} \\
&m \mathcal{E}_{ij} \frac{x_p x_q x^i x^j}{\rho^3}, \:
 m \eta_{pq} \mathcal{E}_{ij} \frac{x^i x^j}{\rho}
\end{split}
\end{displaymath}
with appropriate constants multiplying each term so that the differential
equations are satisfied.
Eq.~(\ref{EBRel2}) and (\ref{EBRel4}) can be used to eliminate the
second and third terms in favor of the remaining two, for the expression
for the $p-q$ spatial components. Substituting into the differential
equations we get simple algebraic equations for the multiplicative
constants. The result is that the first correction to the singular
field has components
\begin{equation}
\begin{split}
h_{\text{S}}^{(2)}{}_{tt} &= 2 m \mathcal{E}_{ij} \frac{x^i x^j}{\rho} \\
h_{\text{S}}^{(2)}{}_{tp} &= -\frac{2}{3} m \epsilon_{pij} 
\mathcal{B}^i{}_k \frac{x^j x^k}{\rho} \\
h_{\text{S}}^{(2)}{}_{pq} &= -4 m \mathcal{E}_{pq} \rho 
  -2 m \eta_{pq} \mathcal{E}_{ij} \frac{x^i x^j}{\rho}. \\
\end{split}
\end{equation}

The second correction is also a symmetric tensor and must obey the
differential equations derived from Eq.~(\ref{LinEinstein2}). Substituting the
components of the second correction results in ten differential equations.
As in the case of the first correction, four of these equations 
involve second derivatives of the four diagonal components. Each one of
the other six differential equations contains the second derivatives
of the six non-diagonal components of the second correction. The
four equations for the diagonal components and the three equations for the 
non-diagonal purely spatial components relate derivatives of those components
to terms of the form
\begin{displaymath}
m \dot{\mathcal{B}}_{..} \frac{x^. x^. x^.}{\rho}, \: 
m \mathcal{E}_{...} \frac{x^. x^. x^.}{\rho}.
\end{displaymath}
The equations for the $t-p$ components relate their derivatives to
terms of the form
\begin{displaymath}
m \dot{\mathcal{E}}_{..} \frac{x^. x^. x^.}{\rho}, \:
m \mathcal{B}_{...} \frac{x^. x^. x^.}{\rho}.
\end{displaymath}
We can predict the form of the solution as previously. The $t-t$ component
must have no free indices. It can be easily deduced that all possible
terms that contain $\dot{\mathcal{B}}$ and satisfy that
requirement are identically equal to zero, 
so the $t-t$ component is expected to contain only the $\mathcal{E}_{ijk}$
tensor and be proportional to
\begin{displaymath}
m \mathcal{E}_{ijk} \frac{x^i x^j x^k}{\rho}.
\end{displaymath}
The $t-p$ components are expected to be a sum of the terms
\begin{displaymath}
\begin{split}
&m \dot{\mathcal{E}}_{pi} x^i \rho, \: m \dot{\mathcal{E}}_{ij}
\frac{x^i x^j x_p}{\rho}, \\
&m \epsilon_{pij} \mathcal{B}^i{}_{kl} \frac{x^j x^k x^l}{\rho},
\end{split}
\end{displaymath}
and the $p-q$ spatial components are expected to be a sum of the terms
\begin{displaymath}
\begin{split}
& m \epsilon_{ij(p} \dot{\mathcal{B}}_{q)}{}^i x^j \rho, \:
  m \epsilon_{ij(p} x_{q)} \dot{\mathcal{B}}^i{}_k \frac{x^j x^k}{\rho} \\
& \eta_{pq} m \mathcal{E}_{ijk} \frac{x^i x^j x^k}{\rho}, \:
  m \mathcal{E}_{pqi} x^i \rho, \: 
  m \mathcal{E}_{ij(p} x_{q)} \frac{x^i x^j}{\rho},
\end{split}
\end{displaymath}
each term multiplied by an appropriate factor so that the differential
equations are satisfied.
The parantheses around two indices denote the usual symmatrization 
convention with respect to
those indices, for example
\begin{displaymath}
\epsilon_{i j (p} x_{q)} x^i x^j = 
\frac{1}{2} (\epsilon_{ijp} x_q + \epsilon_{ijq} x_p) x^i x^j.
\end{displaymath}
Substituting these expressions into the differential equations gives simple
algebraic equations for the multiplicative constants, which can be easily
solved. The second correction is then equal to
\begin{equation}
\begin{split}
h_{\text{S}}^{(3)}{}_{tt} =& \frac{2}{3} m \mathcal{E}_{ijk} 
                        \frac{x^i x^j x^k}{\rho} \\
h_{\text{S}}^{(3)}{}_{tp} =& -\frac{10}{63} m \dot{\mathcal{E}}_{pi}x^i \rho
  -\frac{38}{63} m \dot{\mathcal{E}}_{ij} \frac{x_p x^i x^j}{\rho} \\
 & -\frac{2}{9} m \epsilon_{pij} \mathcal{B}^i{}_{kl} \frac{x^j x^k x^l}{\rho}\\
h_{\text{S}}^{(3)}{}_{pq} =& \frac{10}{7} m \epsilon_{ij(p} 
\dot{\mathcal{B}}_{q)}
  {}^i x^j \rho
  -\frac{10}{21} m \epsilon_{ij(p} x_{q)} \dot{\mathcal{B}}^i{}_k 
    \frac{x^j x^k}{\rho} \\
  & -\frac{2}{3} \eta_{pq} m \mathcal{E}_{ijk} \frac{x^i x^j x^k}{\rho}
    -2 m \mathcal{E}_{pqi}x^i \rho. \\
\end{split}
\end{equation}

The order of the next correction can be predicted by Eq.~(\ref{LinEinstein3})
and it is
\begin{equation}
h_{\text{S} (4)} \sim O(\frac{\rho^3}{\mathcal{R}^4}).
\end{equation}
Evaluating these terms is clearly not necessary for self-force calculations 
because
they will result in zero contribution after the first derivative and
the limit to the location of the particle is taken.

Finally, the singular gravitational field of a point particle of mass m
moving on a geodesic is equal to
\begin{equation}
\begin{split}
h_{\text{S}}{}_{tt} =&  \frac{2m}{\rho} +
  2 m \mathcal{E}_{ij} \frac{x^i x^j}{\rho} 
 + \frac{2}{3} m \mathcal{E}_{ijk}
   \frac{x^i x^j x^k}{\rho}+O(\frac{\rho^3}{\mathcal{R}^4}) \\
h_{\text{S}}{}_{tp} =& -\frac{2}{3} m \epsilon_{pij}
\mathcal{B}^i{}_k \frac{x^j x^k}{\rho} \\
& -\frac{10}{63} m \dot{\mathcal{E}}_{pi}x^i \rho
  -\frac{38}{63} m \dot{\mathcal{E}}_{ij} \frac{x_p x^i x^j}{\rho} \\
 & -\frac{2}{9} m \epsilon_{pij} \mathcal{B}^i{}_{kl} \frac{x^j x^k x^l}{\rho}
  + O(\frac{\rho^3}{\mathcal{R}^4})\\
h_{\text{S}}{}_{pq} =& \eta_{pq} \frac{2m}{\rho} -4 m \mathcal{E}_{pq} \rho
  -2 m \eta_{pq} \mathcal{E}_{ij} \frac{x^i x^j}{\rho} \\
&+ \frac{10}{7} m \epsilon_{ij(p} \dot{\mathcal{B}}_{q)}
  {}^i x^j \rho
  -\frac{10}{21} m \epsilon_{ij(p} x_{q)} \dot{\mathcal{B}}^i{}_k
    \frac{x^j x^k}{\rho} \\
  & -\frac{2}{3} \eta_{pq} m \mathcal{E}_{ijk} \frac{x^i x^j x^k}{\rho}
    -2 m \mathcal{E}_{pqi}x^i \rho + O(\frac{\rho^3}{\mathcal{R}^4}).\\
\end{split}
\label{hSgrav}
\end{equation}
Direct substitution into Eq.~(\ref{HarmonicGauge}) shows that this
expression for the singular field is consistent with
the harmonic gauge condition.

It is clear that, in this case, the second correction 
to the singular gravitational field is not necessary
for the calculation of the self-force on the massive particle. 
After differentiation, that correction
will give terms proportional to the coordinates which, after taking
the limit to the location of the particle, will give zero.

\section{Discussion}
\label{Discussion}

There is a very obvious categorization of the particles mentioned above, based
on the order up to which it is necessary to calculate the singular 
fields in order
to be able to calculate the self-force. For the particles that carry no 
intrinsic spin,
namely the scalar monopole particle, the charged particle and the
non-spinning massive
particle, it is sufficient to calculate the singular fields/potentials up to,
and including, first corrections. For the particles that do carry 
intrinsic spin,
namely the scalar dipole particle and the particle carrying an electric or a
magnetic dipole moment, it is necessary to
include the second correction to the singular fields, in
order to be able to calculate the self-force.

That can be explained quite easily by considering the fields generated
by each particle and we use here the simple example of
the scalar fields. Let us compare the scalar monopole field at a point 
$\vec{r}$
\begin{equation}
\psi(\vec{r}) = \frac{q}{|\vec{r}-\vec{r_0}|}
\end{equation}
of a scalar charge $q$ at point $\vec{r_o}$ to the dipole field at a
point $\vec{r}$
\begin{equation}
\psi(\vec{r}) = \frac{K_i (x^i-x_o{}^i) }{|\vec{r}-\vec{r_0}|^3}
\end{equation}
of a dipole moment $\vec{K}$ at point $\vec{r_o}$. 
The monopole field is less singular at the location of the particle than
the dipole field is. The former falls off as $1/(|\vec{r}-\vec{r_0}|)$ while
the latter falls off as $1/(|\vec{r}-\vec{r_0}|)^2$, in the limit $\vec{r} \to 
\vec{r_0}$. Differently said, the field of the non-spinning particle
is ``less singular'' in the limit $\vec{r} \to \vec{r_0}$ than the field
of the spinning partilce is.

\section{Future work}
\label{future}

The author and collaborators are currently
working on a self-force calculation for
a particle of mass $m$ in the vicinity of a Schwarzschild black hole, as that
calculation is described in \cite{KFW2006}. The ultimate goal is to
generalize the calculation to the more interesting (for LISA data
analysis) case of a particle
moving in the vicinity of a Kerr black hole. Knowledge of either the singular 
gravitational field given in Eq.~(\ref{hSgrav}) or of the direct contribution
to the retarded gravitational field is necessary for those calculations.

The author is also working on calculating the part of the self-force 
that is due to the 
spin of a particle. The calculation of the singular gravitational field
of a spinning particle will be presented in 
a future paper. Even though we intend to
follow the general method described here, that calculation is more complicated
than those described in this paper. At that time we will also present the
spherical harmonic decomposition of the singular fields, which is necessary
for the self-force calculation as was first described in \cite{bo0} and
outlined in Sec.~\ref{Calc} in this paper.
We will first carry out the spherical harmonic decomposition for the 
toy-problem
of the scalar dipole field given in Eq.~(\ref{scalardipole}) to try and
identify any issues that arise from the presence of intrinsic spin
and then we will move on to performing the same calculation for 
the gravitational field of the spinning particle.

\begin{acknowledgments}
I wish to thank Steve Detweiler for his guidance in completing
this work and for his very useful suggestions on this manuscript. I
also wish to thank Bernard Whiting for useful discussions. This
work was supported in part by the Institute for Fundamental
Theory at the University of Florida and in part by NSF PHY-0200852
at the University of Wisconsin - Milwaukee.
\end{acknowledgments}

\bibliography{References}

\begin{thebibliography}{27}
\expandafter\ifx\csname natexlab\endcsname\relax\def\natexlab#1{#1}\fi
\expandafter\ifx\csname bibnamefont\endcsname\relax
  \def\bibnamefont#1{#1}\fi
\expandafter\ifx\csname bibfnamefont\endcsname\relax
  \def\bibfnamefont#1{#1}\fi
\expandafter\ifx\csname citenamefont\endcsname\relax
  \def\citenamefont#1{#1}\fi
\expandafter\ifx\csname url\endcsname\relax
  \def\url#1{\texttt{#1}}\fi
\expandafter\ifx\csname urlprefix\endcsname\relax\def\urlprefix{URL }\fi
\providecommand{\bibinfo}[2]{#2}
\providecommand{\eprint}[2][]{\url{#2}}

\bibitem[{LIS(http://lisa.jpl.nasa.gov/)}]{LISAreference}
\bibinfo{journal}{LISA website available}
  (\bibinfo{year}{http://lisa.jpl.nasa.gov/}).

\bibitem[{\citenamefont{DeWitt and Brehme}(1960)}]{dewbre}
\bibinfo{author}{\bibfnamefont{B.~S.} \bibnamefont{DeWitt}} \bibnamefont{and}
  \bibinfo{author}{\bibfnamefont{R.~W.} \bibnamefont{Brehme}},
  \bibinfo{journal}{Annals of Physics} \textbf{\bibinfo{volume}{9}},
  \bibinfo{pages}{220} (\bibinfo{year}{1960}).

\bibitem[{\citenamefont{Mino et~al.}(1997)\citenamefont{Mino, Sasaki, and
  Tanaka}}]{mst}
\bibinfo{author}{\bibfnamefont{Y.}~\bibnamefont{Mino}},
  \bibinfo{author}{\bibfnamefont{M.}~\bibnamefont{Sasaki}}, \bibnamefont{and}
  \bibinfo{author}{\bibfnamefont{T.}~\bibnamefont{Tanaka}},
  \bibinfo{journal}{Phys. Rev. D} \textbf{\bibinfo{volume}{55}},
  \bibinfo{pages}{3457} (\bibinfo{year}{1997}).

\bibitem[{\citenamefont{Quinn}(2000)}]{q00}
\bibinfo{author}{\bibfnamefont{T.~C.} \bibnamefont{Quinn}},
  \bibinfo{journal}{Phys. Rev. D} \textbf{\bibinfo{volume}{62}},
  \bibinfo{pages}{064029} (\bibinfo{year}{2000}).

\bibitem[{\citenamefont{Quinn and Wald}(1997)}]{qw97}
\bibinfo{author}{\bibfnamefont{T.~C.} \bibnamefont{Quinn}} \bibnamefont{and}
  \bibinfo{author}{\bibfnamefont{R.~M.} \bibnamefont{Wald}},
  \bibinfo{journal}{Phys. Rev. D} \textbf{\bibinfo{volume}{56}},
  \bibinfo{pages}{3381} (\bibinfo{year}{1997}).

\bibitem[{\citenamefont{Mino et~al.}(2003)\citenamefont{Mino, Nakano, and
  Sasaki}}]{mns}
\bibinfo{author}{\bibfnamefont{Y.}~\bibnamefont{Mino}},
  \bibinfo{author}{\bibfnamefont{H.}~\bibnamefont{Nakano}}, \bibnamefont{and}
  \bibinfo{author}{\bibfnamefont{M.}~\bibnamefont{Sasaki}},
  \bibinfo{journal}{Prog. Theor. Phys.} \textbf{\bibinfo{volume}{108}},
  \bibinfo{pages}{1039} (\bibinfo{year}{2003}).

\bibitem[{\citenamefont{Barack and Ori}(2000)}]{bo0}
\bibinfo{author}{\bibfnamefont{L.}~\bibnamefont{Barack}} \bibnamefont{and}
  \bibinfo{author}{\bibfnamefont{A.}~\bibnamefont{Ori}},
  \bibinfo{journal}{Phys. Rev. D} \textbf{\bibinfo{volume}{61}},
  \bibinfo{pages}{061502(R)} (\bibinfo{year}{2000}).

\bibitem[{\citenamefont{Barack and Ori}(2003)}]{bo2}
\bibinfo{author}{\bibfnamefont{L.}~\bibnamefont{Barack}} \bibnamefont{and}
  \bibinfo{author}{\bibfnamefont{A.}~\bibnamefont{Ori}},
  \bibinfo{journal}{Phys. Rev. D} \textbf{\bibinfo{volume}{67}},
  \bibinfo{pages}{024029} (\bibinfo{year}{2003}).

\bibitem[{\citenamefont{Barack and Ori}(2002)}]{bo1}
\bibinfo{author}{\bibfnamefont{L.}~\bibnamefont{Barack}} \bibnamefont{and}
  \bibinfo{author}{\bibfnamefont{A.}~\bibnamefont{Ori}},
  \bibinfo{journal}{Phys. Rev. D} \textbf{\bibinfo{volume}{66}},
  \bibinfo{pages}{084022} (\bibinfo{year}{2002}).

\bibitem[{\citenamefont{Burko}(2000)}]{burko}
\bibinfo{author}{\bibfnamefont{L.~M.} \bibnamefont{Burko}},
  \bibinfo{journal}{Phys. Rev. Lett.} \textbf{\bibinfo{volume}{84}},
  \bibinfo{pages}{4529} (\bibinfo{year}{2000}).

\bibitem[{\citenamefont{Barack and Burko}(2000)}]{BarBur}
\bibinfo{author}{\bibfnamefont{L.}~\bibnamefont{Barack}} \bibnamefont{and}
  \bibinfo{author}{\bibfnamefont{L.~M.} \bibnamefont{Burko}},
  \bibinfo{journal}{Phys. Rev. D} \textbf{\bibinfo{volume}{62}},
  \bibinfo{pages}{084040} (\bibinfo{year}{2000}).

\bibitem[{\citenamefont{Barack}(2000)}]{barack1}
\bibinfo{author}{\bibfnamefont{L.}~\bibnamefont{Barack}},
  \bibinfo{journal}{Phys. Rev. D} \textbf{\bibinfo{volume}{62}},
  \bibinfo{pages}{084027} (\bibinfo{year}{2000}).

\bibitem[{\citenamefont{Barack}(2001)}]{barack2}
\bibinfo{author}{\bibfnamefont{L.}~\bibnamefont{Barack}},
  \bibinfo{journal}{Phys. Rev. D} \textbf{\bibinfo{volume}{64}},
  \bibinfo{pages}{084021} (\bibinfo{year}{2001}).

\bibitem[{\citenamefont{Barack et~al.}(2002)\citenamefont{Barack, Mino, Nakano,
  Ori, and Sasaki}}]{BMetal}
\bibinfo{author}{\bibfnamefont{L.}~\bibnamefont{Barack}},
  \bibinfo{author}{\bibfnamefont{Y.}~\bibnamefont{Mino}},
  \bibinfo{author}{\bibfnamefont{H.}~\bibnamefont{Nakano}},
  \bibinfo{author}{\bibfnamefont{A.}~\bibnamefont{Ori}}, \bibnamefont{and}
  \bibinfo{author}{\bibfnamefont{M.}~\bibnamefont{Sasaki}},
  \bibinfo{journal}{Phys. Rev. Lett.} \textbf{\bibinfo{volume}{88}},
  \bibinfo{pages}{091101} (\bibinfo{year}{2002}).

\bibitem[{\citenamefont{Barack and Lousto}(2002)}]{BarackLousto02}
\bibinfo{author}{\bibfnamefont{L.}~\bibnamefont{Barack}} \bibnamefont{and}
  \bibinfo{author}{\bibfnamefont{C.~O.} \bibnamefont{Lousto}},
  \bibinfo{journal}{PRD} \textbf{\bibinfo{volume}{66}},
  \bibinfo{pages}{061502(R)} (\bibinfo{year}{2002}), \eprint{gr-qc/0205043}.

\bibitem[{\citenamefont{Detweiler and Whiting}(2003)}]{detwhi02}
\bibinfo{author}{\bibfnamefont{S.}~\bibnamefont{Detweiler}} \bibnamefont{and}
  \bibinfo{author}{\bibfnamefont{B.~F.} \bibnamefont{Whiting}},
  \bibinfo{journal}{Phys. Rev. D} \textbf{\bibinfo{volume}{67}},
  \bibinfo{pages}{024025} (\bibinfo{year}{2003}).

\bibitem[{\citenamefont{Poisson}(2004)}]{PoissonsReview}
\bibinfo{author}{\bibfnamefont{E.}~\bibnamefont{Poisson}},
  \bibinfo{journal}{Living Rev. Relativity} \textbf{\bibinfo{volume}{7}}
  (\bibinfo{year}{2004}), \eprint{gr-qc/0306052}.

\bibitem[{\citenamefont{Detweiler et~al.}(2003)\citenamefont{Detweiler,
  Messaritaki, and Whiting}}]{detmeswhi}
\bibinfo{author}{\bibfnamefont{S.}~\bibnamefont{Detweiler}},
  \bibinfo{author}{\bibfnamefont{E.}~\bibnamefont{Messaritaki}},
  \bibnamefont{and} \bibinfo{author}{\bibfnamefont{B.~F.}
  \bibnamefont{Whiting}}, \bibinfo{journal}{Phys. Rev. D}
  \textbf{\bibinfo{volume}{67}}, \bibinfo{pages}{104016}
  (\bibinfo{year}{2003}).

\bibitem[{\citenamefont{Diaz et~al.}(2004)\citenamefont{Diaz, Messaritaki,
  Whiting, and Detweiler}}]{diazmesswhidet}
\bibinfo{author}{\bibfnamefont{L.~M.} \bibnamefont{Diaz}},
  \bibinfo{author}{\bibfnamefont{E.}~\bibnamefont{Messaritaki}},
  \bibinfo{author}{\bibfnamefont{B.~F.} \bibnamefont{Whiting}},
  \bibnamefont{and}
  \bibinfo{author}{\bibfnamefont{S.}~\bibnamefont{Detweiler}},
  \bibinfo{journal}{Phys. Rev. D} \textbf{\bibinfo{volume}{70}},
  \bibinfo{pages}{124018} (\bibinfo{year}{2004}), \eprint{gr-qc/0410011}.

\bibitem[{\citenamefont{Keidl et~al.}(2006)\citenamefont{Keidl, Friedman, and
  Wiseman}}]{KFW2006}
\bibinfo{author}{\bibfnamefont{T.~S.} \bibnamefont{Keidl}},
  \bibinfo{author}{\bibfnamefont{J.~L.} \bibnamefont{Friedman}},
  \bibnamefont{and} \bibinfo{author}{\bibfnamefont{A.~G.}
  \bibnamefont{Wiseman}} (\bibinfo{year}{2006}), \eprint{gr-qc/0611072}.

\bibitem[{\citenamefont{Thorne and Hartle}(1985)}]{TH}
\bibinfo{author}{\bibfnamefont{K.~S.} \bibnamefont{Thorne}} \bibnamefont{and}
  \bibinfo{author}{\bibfnamefont{J.~B.} \bibnamefont{Hartle}},
  \bibinfo{journal}{Phys. Rev. D} \textbf{\bibinfo{volume}{31}},
  \bibinfo{pages}{1815} (\bibinfo{year}{1985}).

\bibitem[{\citenamefont{Zhang}(1986)}]{Z}
\bibinfo{author}{\bibfnamefont{X.~H.} \bibnamefont{Zhang}},
  \bibinfo{journal}{Phys. Rev. D} \textbf{\bibinfo{volume}{34}},
  \bibinfo{pages}{991} (\bibinfo{year}{1986}).

\bibitem[{\citenamefont{Misner et~al.}(1973)\citenamefont{Misner, Thorne, and
  Wheeler}}]{mtw}
\bibinfo{author}{\bibfnamefont{C.~W.} \bibnamefont{Misner}},
  \bibinfo{author}{\bibfnamefont{K.~S.} \bibnamefont{Thorne}},
  \bibnamefont{and} \bibinfo{author}{\bibfnamefont{J.~A.}
  \bibnamefont{Wheeler}}, \textbf{\bibinfo{volume}{(Freeman, San Fransisco)}}
  (\bibinfo{year}{1973}).

\bibitem[{\citenamefont{Detweiler}(2005)}]{Det2005}
\bibinfo{author}{\bibfnamefont{S.~L.} \bibnamefont{Detweiler}},
  \bibinfo{journal}{Class. Quantum Grav.} \textbf{\bibinfo{volume}{22}},
  \bibinfo{pages}{S681} (\bibinfo{year}{2005}).

\bibitem[{grt(1992)}]{grtensor}
\bibinfo{journal}{GRTENSOR software website available:}
  (\bibinfo{year}{1992}), \eprint{http://grtensor.phy.queensu.ca/}.

\bibitem[{\citenamefont{Hadamard}(1923)}]{hadamard}
\bibinfo{author}{\bibfnamefont{J.}~\bibnamefont{Hadamard}},
  \textbf{\bibinfo{volume}{(Yale University Press, New Haven)}}
  (\bibinfo{year}{1923}).

\bibitem[{\citenamefont{Messaritaki}(2003)}]{mythesis}
\bibinfo{author}{\bibfnamefont{E.}~\bibnamefont{Messaritaki}},
  \bibinfo{journal}{Ph.D. Dissertation}  (\bibinfo{year}{2003}),
  \eprint{gr-qc/0308047}.

\end{thebibliography}

\end{document}